\documentclass[prb,preprint,showpacs]{revtex4}
\usepackage{graphicx}
\begin{document}
\title{Controlling the gap of fullerene microcrystals by applying pressure: 
the role of many-body effects}

\author{Murilo L. Tiago}
\affiliation{Oak Ridge National Laboratory, Oak Ridge, TN, 37831} 
\author{Fernando A. Reboredo}
\affiliation{Oak Ridge National Laboratory, Oak Ridge, TN, 37831}
\date{\today}

\begin{abstract}
We studied theoretically the optical properties of C$_{60}$ fullerene
microcrystals as a function of hydrostatic pressure with first-principles 
 many-body theories. 
Calculations of the electronic properties were done in the GW approximation. We computed
electronic excited states in the crystal by diagonalizing the
Bethe-Salpeter equation (BSE). Our results confirmed the existence of
bound excitons in the crystal. Both the electronic gap and optical gap
decrease continuously and non-linearly as pressure of up to 6~GPa is
applied. As a result, the absorption spectrum shows strong redshift. We also
obtained that ``negative'' pressure shows the opposite
behavior: the gaps increase and the optical spectrum shifts toward the
blue end of the spectrum. Negative pressure can be realized by adding
cubane (C$_8$H$_8$) or other molecules with similar size to the interstitials of the
microcrystal. For the moderate lattice distortions studied
here, we found that the optical
properties of fullerene microcrystals with intercalated cubane are
similar to the ones of an expanded undoped microcrystal. Based on these findings, we
propose doped C$_{60}$ as active element in piezo-optical devices.
\end{abstract}

\maketitle

\section{Introduction} \label{introduction}

Since its discovery, C$_{60}$ \cite{KrotoHOCS85} has been
characterized as the most stable member in the series of fullerenes,
which are pure-carbon molecules with the shape of spherical shells
\cite{DresselhausDE96}. C$_{60}$ can be produced economically and in
abundance \cite{KratschmerLFH90}. Its chemical bonds have strong
$sp^2$ character, making the shell very stiff and at the same time
free of dangling bonds. 
Together with clathrates and other carbon-based materials, fullerenes
are being actively investigated as building blocks for novel materials
with unusual mechanical properties \cite{San-Miguel06,BlaseGSM04}.
Fullerenes has remarkable properties, for example:
C$_{60}$ doped with alkali and alkali-earth atoms is superconductor
\cite{KortanKGGRFTH92,HebardRHMGPRK91}; it has been claimed that C$_{60}$ can be heavily hydrogenated
with up to 36 hydrogen atoms per molecule, making it a promising
material for hydrogen storage \cite{MeletovK05}.

In pure form, C$_{60}$ crystallizes in a molecular solid (fullerite),
bound by weak forces between molecules. Its phase diagram is very
rich, with a low-temperature face-centered cubic (FCC) phase,
amorphous phases at intermediate temperature, and a diamond-like phase
at high temperature. The FCC phase has rotational disorder and it
is stable under pressure in excess of 15~GPa at room temperature
\cite{Sundqvist99}. Its energy gap is in the visible range, around
2~eV \cite{DresselhausDE96,MeletovD98}. The softness and stability of
fullerite could make it a good candidate for piezo-optical devices, where
external pressure is applied reversibly and it modifies the optical
response of the material. In addition, when fullerite is heavily doped with
molecules of appropriate size, it behaves as if it is under ``negative
pressure'' \cite{PekkerKOBKBJJBKBKTF05}. To that end, it is important to characterize
the pressure dependence of the optical properties of this
material. Extensive experimental work has been done on this direction
\cite{PoloniFFPTMCPS08,Duclos91,NunezMHBP95,SnokeSM93,MeletovD98,MosharyCSBDVB92,KozlovY95}. A
review of the literature can be found in Reference
\onlinecite{Sundqvist99}. Theoretical analyses are not so extensive,
mostly concentrated on characterization at zero pressure
\cite{ShirleyL93,ShirleyBL96,HartmannZMSBLFZ95}. In order to fill this
vacuum, we present a systematic analysis of the optical properties of
fullerite at pressures ranging from zero to 6~GPa. We also investigate
the optical response of the crystal at ``negative'' pressure, which
can be realized in laboratory, for instance, by doping fullerite with
weakly interacting molecules such as cubane
(C$_8$H$_8$). Numerical accuracy is essential, which is why we use
state-of-the-art theories, namely many-body Green's function theories
based on the GW-BSE approach.

This article is organized as follows: we outline the theoretical
framework in section \ref{theory}. That section is followed by a
discussion of results at zero pressure in section
\ref{results_equilibrium}, results at finite hydrostatic pressure in
section \ref{results_pressure} and finally a description of the
cubane-fullerene compound, which resembles fullerite with ``negative
pressure'', in section \ref{results_cubane}. Finally, we conclude with
some perspectives of future applications and a summary.

\section{Theory} \label{theory}
The underlying electronic structure of fullerene is determined using
density-functional theory (DFT) \cite{Martin}. We use a plane-wave
basis to solve the Kohn-Sham equations with cut-off in the kinetic
energy of 50~Ry. Interactions involving valence electrons and core
electrons are taken into account using norm-conserving
pseudopotentials of the Troullier-Martins type \cite{Martin}. We use
the Perdew-Burke-Ernzerhof functional \cite{PerdewBE96} (PBE) for exchange
and correlation, based on the generalized-gradient approximation
(GGA). It is well-known that DFT in the local-density approximation
(LDA) or the PBE severely underestimates electronic gaps in general,
making it unsuitable for detailed studies of optical properties of
electronic systems. Accurate bandwidths and electronic energy gaps are calculated
in a many-body framework within the GW approximation
\cite{HybertsenL86}. In that approximation, the electron self-energy
is computed by summing up Feynman diagrams to lowest order in the
screened Coulomb interaction. At lowest order, the self-energy becomes
a product between the one-electron Green's function $G$ and the
screened Coulomb interaction $W$, hence the name.
We ignore vertex diagrams and we assume
that Kohn-Sham eigenvalues and eigenvectors give a good approximation
to the Green's function. Formally, the self-energy in space-energy
representation is written as
\begin{equation}
\Sigma ({\bf r},{\bf r}^\prime;E) = {i \over 2 \pi} \int {\rm
  d}E^\prime e^{-i0^+E^\prime} G_0({\bf r},{\bf r}^\prime;E -
  E^\prime) W_0({\bf r},{\bf r}^\prime;E^\prime) \; \; ,
\label{eq.gw0}
\end{equation}
where $G_0$ denotes the DFT-PBE Green's function and $W_0$ is the screened
Coulomb interaction, related to the random phase approximation (RPA)
dielectric function by:
\begin{equation}
W_0 ({\bf r},{\bf r}^\prime;E) = \int {\rm
  d}{\bf r}^{\prime\prime} \epsilon^{-1}({\bf r},{\bf
  r}^{\prime\prime};E) {q_e^2 \over |{\bf r}^{\prime\prime} - {\bf
  r}^\prime|} \; \; .
\label{eq.rpa}
\end{equation}
In the present formulation, the dielectric function is expanded in a
basis of plane waves with cut-off 9.5~Ry. Its energy dependence is
described by a generalized plasmon pole model
\cite{HybertsenL86}. After the self-energy is computed, we diagonalize
the quasi-particle Hamiltonian, $H = H_{PBE} + \Sigma - V_{xc}$
\cite{HybertsenL86,AulburJW00}. Eigenvalues of that Hamiltonian
provide the electronic band structure of the real material. This
formulation is one of the simplest {\em ab initio} formulations of the
GW approximation. Extensive applications of this formulation to a wide
class of carbon-based materials have shown it to predict electronic
band gaps with an accuracy of 0.1 to 0.2~eV
\cite{AulburJW00,OnidaRR02}. In the specific case of fullerite, the first calculation of
electronic gap within the GW approximation was consistent with
direct/inverse photoemission spectra \cite{ShirleyL93}. Owing to the
fact that hydrostatic pressure on fullerite microcrystals does not
affect their electronic properties besides an increase in
intermolecular interactions, we expect our theoretical methodology to
be equally reliable in describing the pressure dependence of the
electronic gap. Technical details about the theory can be found in
review articles \cite{AulburJW00,OnidaRR02}.

The electronic band structure often does not give access to optical
spectra because, after electron-hole pairs are excited, they interact
and produce bound excitons, with energy lower than the electronic gap
\cite{OnidaRR02,RohlfingL00}. We describe the dynamics of excitons by
diagonalizing the Bethe-Salpeter equation (BSE) for electrons and
holes. The BSE is an equation for the two-particle Green's
function. Written as an eigenvalue equation, its solution gives the
energy of optical excitations in the material:

\begin{equation}
(E_c - E_v) A^s_{vc} + \sum_{v^\prime c^\prime} K^{vc}_{v^\prime
    c^\prime} A^s_{v^\prime c^\prime} = \Omega^s A^s_{vc} \; \; .
\label{eq.bse}
\end{equation}
where $\Omega^s$ is the excitation energy of optical modes, indexed by
$s$, and $A^s_{vc}$ are the corresponding
eigenvectors. $K^{vc}_{v^\prime c^\prime}$ is the electron-hole
interaction kernel, written in the basis of pair transitions.

In the absence of electron-hole interactions ($K = 0$), each
excitation energy is simply the difference between quasi-particle
energies of electrons ($E_c$) and holes ($E_v$). The kernel $K$ adds
two types of interactions: an electrostatic interaction mediated again
by $W_0$; and a repulsive exchange interaction between electron and
hole, which is related to the fact that they can annihilate each other\cite{FetterW}. We
follow the standard procedure to build and solve the BSE. We ignore
the energy dependence of the interaction kernel and we assume the
Tamm-Dancoff approximation \cite{FetterW} when computing $K$. Both approximations
have been used extensively and they were shown to simplify
considerably the numerical complexity, with little impact on numerical
accuracy. This methodology has been presented in great detail
elsewhere \cite{OnidaRR02,RohlfingL00,TiagoC06}.

Although numerically expensive, the GW-BSE theory has been remarkably
successful in predicting electronic and optical properties of real
materials without the need for phenomenological parameters\cite{AulburJW00,OnidaRR02}. All the
approximations involved, such as the plasmon pole model and the
non-self-consistent assumption, are unambiguously defined. In
addition, {\em sp}-bonded systems such as carbon-based nanostructures
seem to be the ideal materials for this theory, owing to the fact that
they have very weak correlation effects \cite{OnidaRR02,SpataruICL05,TiagoKHR08}.

\section{Fullerite at zero pressure} \label{results_equilibrium}

The phase diagram of C$_{60}$ is extremely complex. At zero pressure
and temperature, it crystallizes in a structure where the orientation
of molecules is random but the molecules form a face-centered cubic
(FCC) structure with lattice parameter around 14.2~\AA
\cite{Sundqvist99}. At room temperature and under mechanical pressure
of 8~GPa, the crystals were found to polymerize in several phases,
with substantial distortion of the cage
\cite{NunezMHBP95,MosharyCSBDVB92}. Since we are primarily concerned
with hydrostatic pressure, we do not consider anisotropic pressure in
this article. With increasing hydrostatic pressure, the
lattice parameter decreases continuously according to Vinet equation
of state \cite{Duclos91}. The energy threshold of optical transmission
also decreases \cite{SnokeSM93}, following the reduction in lattice
parameter. Other absorption edges are also known to redshift with
applied pressure \cite{MeletovD98}.

In our calculations, we apply pressure indirectly by fixing the
lattice parameter and using Vinet equation to map lattice parameter
into hydrostatic pressure \cite{Duclos91}:
\begin{equation}
p (a) = 3 \kappa_0 {1 - x \over x^2 } \exp{ \left[ {3 \over 2} (
    \kappa_0^\prime - 1 ) ( 1 - x ) \right] } \; \; ,
\label{eq.vinet}
\end{equation}
using a bulk modulus $\kappa_0 = 18.1 \pm 1.8$ GPa and its pressure
derivative $\kappa^\prime_0 = 5.7 \pm 0.6$ \cite{Duclos91}. The parameter $x$ is the ratio
between lattice parameters, $x = a / a_0$. The lattice is built in the
$Pa{\overline{3}}$ (= $cP12$) structure, with one molecule per
periodic cell.

\begin{figure}
\includegraphics[width=16cm]{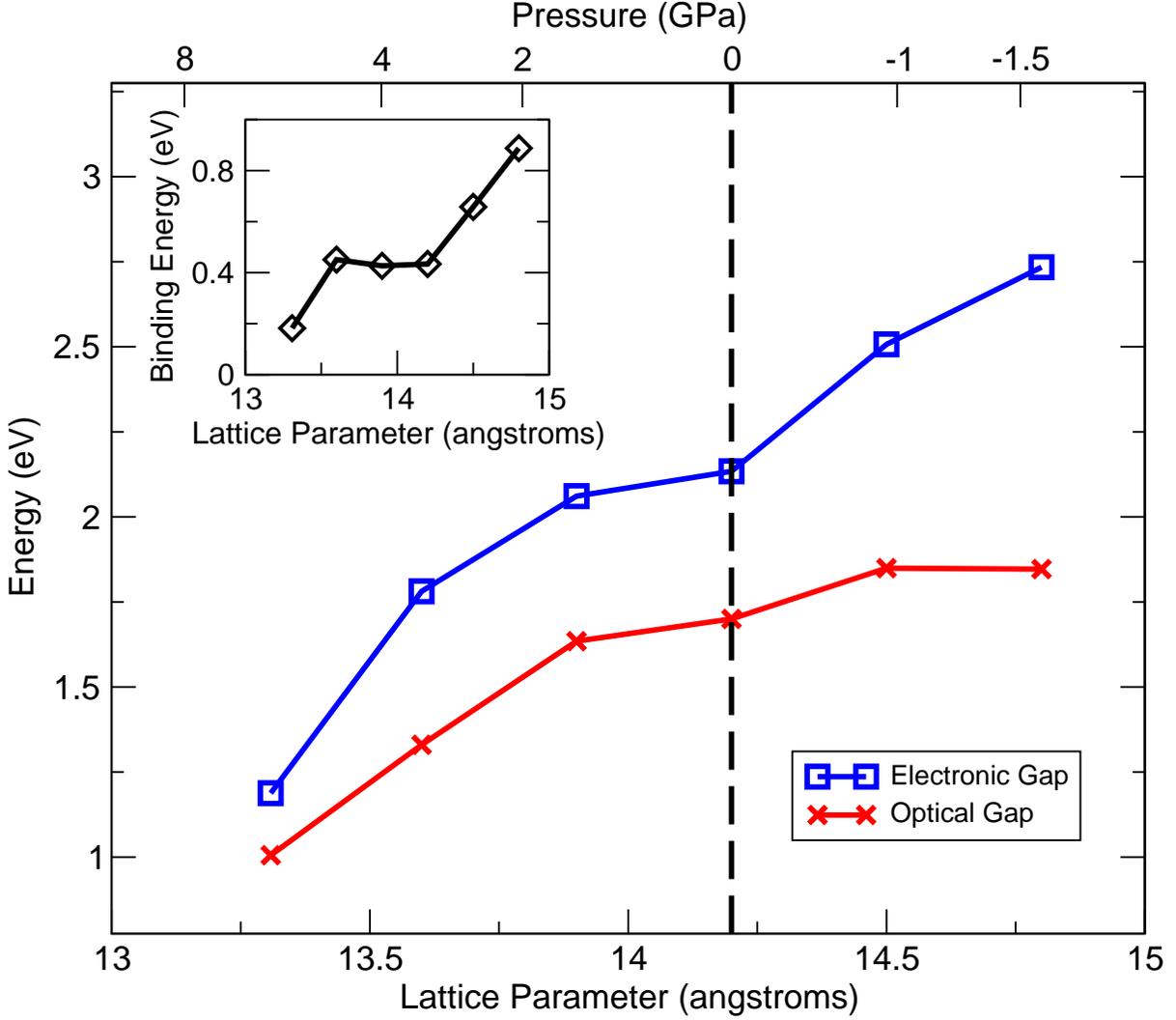}
\caption{(Color online) Electronic gap (squares) and optical gap (crosses) calculated
for fullerite as functions of lattice parameter. The equivalent
hydrostatic pressure was obtained using Vinet equation, Equation
\ref{eq.vinet}. The electronic gap was calculated within the GW
approximation. The optical gap shown is the energy of the first
excitation energy obtained from the BSE. It is a lower bound to the
experimental optical gap since the first excitation has very low oscillator
strength (see text). The inset shows the maximum exciton binding
energy, {\em i.e.} difference between the electronic and optical
gaps.}
\label{fig.gap}
\end{figure}

Figure \ref{fig.gap} shows the calculated electronic and optical gaps
for several choices of lattice parameter. At zero pressure,
we obtain an electronic gap of 2.1~eV, in full agreement
with previous work and compatible with photoemission and inverse
photoemission data \cite{ShirleyL93}. Our DFT-PBE gap is 1.2~eV. The
minimum gap is direct, around the crystallographic $X$ point.

The C$_{60}$ molecule has icosahedral symmetry, belonging to the I$_h$
point group \cite{Cotton}. Owing to its high symmetry, most molecular
orbitals cluster in degenerate multiplets. The three highest occupied
multiplets belong to symmetry representations denoted as H$_u$, G$_g$, and H$_g$
(ordered from highest energy to lowest energy), with degeneracies 5, 4
and 5 respectively. The lowest unoccupied
multiplet in molecular C$_{60}$ has symmetry T$_{1u}$, followed by a
T$_{1g}$ multiplet, both with degeneracy 3.

In fullerite, each molecular multiplet originates a set of quasi-degenerate bands.
The wavefunctions retain most of the shape of the molecular orbitals, so that they can
still be labeled by symmetry representations of the molecular orbitals. The
bandwidth of the $H_u$ quintuplet, at the top of the valence bands, is
0.5~eV. The next multiplet is a $T_{1u}$ triplet, with approximately the same
bandwidth. Within the GW theory, these bandwidths are slightly larger than the ones calculated with DFT-PBE.
There are two major differences between band structures
predicted with GW and DFT-PBE: (1) widening of the electronic gap, and (2)
small stretch of bands according to the expressions:

\begin{eqnarray}
&& E_{GW}^{val.} = E_{PBE}^{val.} \times 1.2 + 0.6 {\rm \; \; eV} \; \;
  \nonumber \\
&& E_{GW}^{cond.} = E_{PBE}^{cond.} \times
1.2  + 1.25 {\rm \; \; eV} \; \; ,
\label{eq.scissors}
\end{eqnarray}
respectively for valence and conduction bands. In the equation above,
the energies $E_{GW}$ and
$E_{PBE}$ are given with respect to the DFT-PBE valence band maximum.

We determine the optical gap as the minimum excitation energy obtained after
diagonalizing the BSE. This gap at equilibrium lattice constant is
calculated to be 1.7~eV. The oscillator strength associated to this
excitation is very weak, owing to a molecular selection rule that prevents optical
absorption from the $H_u$ to the $T_{1u}$ multiplets. Significant
absorption is found around 2.2~eV, corresponding to $H_u$-$T_{1g}$
transitions, as shown on Figure \ref{fig.spectrum}. The measured
transmission edge is 1.9~eV \cite{SnokeSM93}. This is compatible with
our calculated results, considering the difficulties in determining
the onset of absorption experimentally and the orientational disorder in the lattice, which is not
included in our calculations. One of the earliest measurements of
absorption spectra of crystalline C$_{60}$ identified peaks at 2.0~eV,
2.7~eV and 3.5~eV. Our calculations show peaks at 2.2~eV and 3.6~eV.

The inset of Figure \ref{fig.gap} shows the maximum exciton binding
energy, defined as the difference between electronic gap and optical
gap. The binding energy is high: around 0.4~eV. As discussed
above, it arises primarily from the Coulomb attraction between
electrons and holes, which is large compared with other solids because of a weak dielectric screening. The first bound exciton has well-defined Frenkel
character, which is compatible with the fact that its binding energy
is close to the electron and hole bandwidths.

\begin{figure}
\includegraphics[width=16cm]{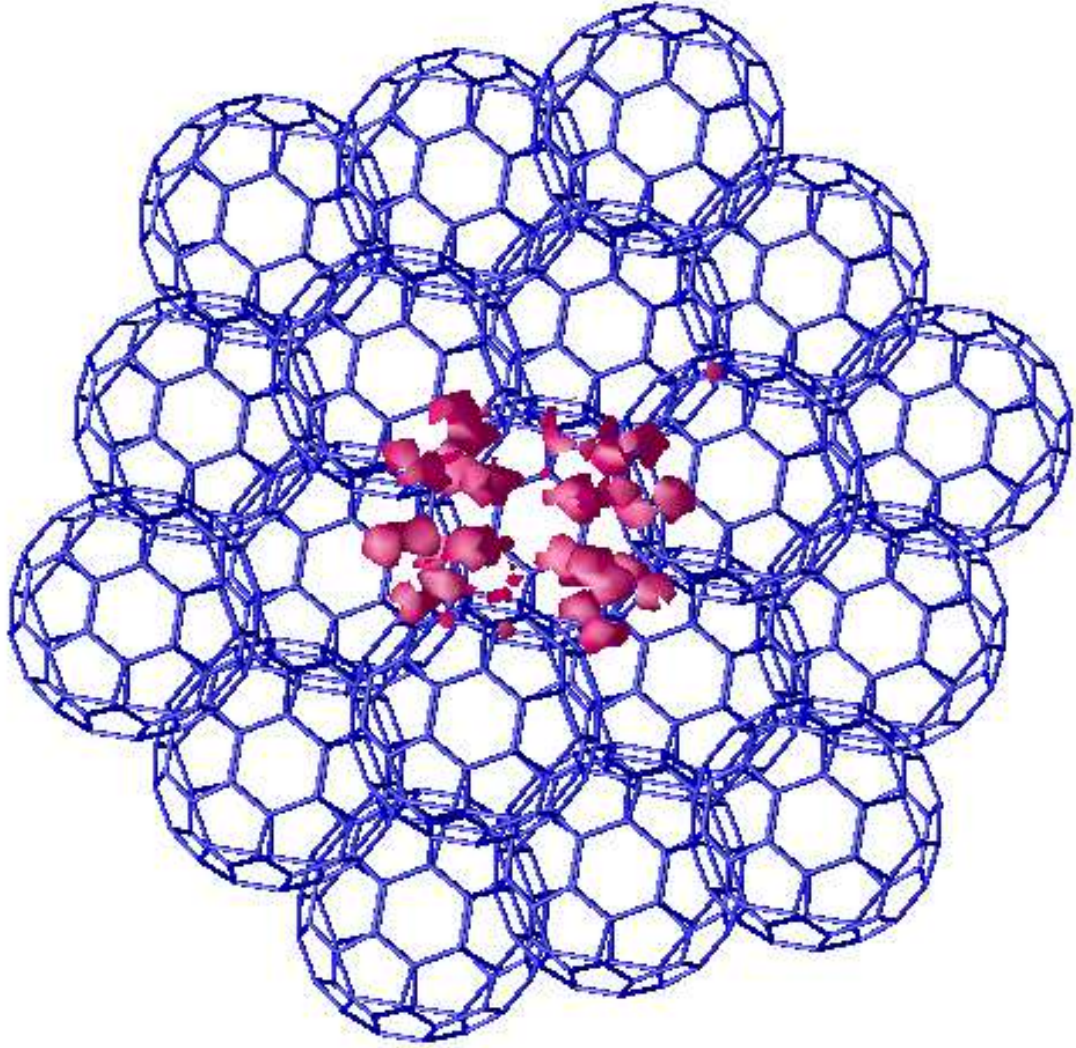}
\caption{(Color online) 
Isocontour plot of the electron probability distribution function 
of the first bound exciton
provided that  the hole is at a position where the
highest occupied molecular orbital has maximum amplitude.
The plot corresponds to fullerite at zero pressure. 
Only molecules up to second neighbor from the hole site 
are depicted in the figure. The isocontour shown corresponds to a value 10\% lower
than the maximum value of the probability distribution. This exciton
is composed primarily by transitions in the $H_u \to T_{1u}$ multiplet.
}
\label{fig.exciton}
\end{figure}

Figure \ref{fig.exciton} depicts the probability
distribution of the electron given that the hole is fixed on the
surface of the central molecule. There are sharp maximums of
probability on the central molecule, with more diffuse features in the
neighbor molecules.

In order to quantify the exciton radius, we have
computed the integrated electron-hole probability and listed it on
the third column of 
Table \ref{table.probability}. For the first bound exciton, the
probability of locating electron and hole on the same molecule is
62\%. The probability of locating the electron on any of the nearest neighbors
molecule relative to the hole site in substantially smaller (30\%),
decreasing then to 2\% if the electron is on any second nearest
neighbor. Excitons with lower binding energy (and higher excitation
energy) have more pronounced charge-transfer character, with the
probability at nearest neighbor higher than the probability at the
hole site.

In order to address the validity of our calculations, based on the
$Pa{\overline{3}}$ lattice, with respect to the real, glassy crystal, we repeated
the zero-pressure calculations with five different orientations of the
molecule. As a result, the electronic gap fluctuated from 2.0 to
2.2~eV. That establishes an uncertainty in the determination of energy
gaps arising from orientational disorder of the molecules. We find
that orientational disorder affects similarly the electronic and
optical gaps. The exciton binding energy fluctuates by less than 0.1~eV
upon rotation of the molecular unit. Fine features in the absorption
spectrum and in the density of states are smoothed out by molecular disorder while the
broader features (energy position and width of major peaks) are very
robust.

\section{Fullerite at hydrostatic pressure} \label{results_pressure}

\begin{figure}
\includegraphics[width=16cm]{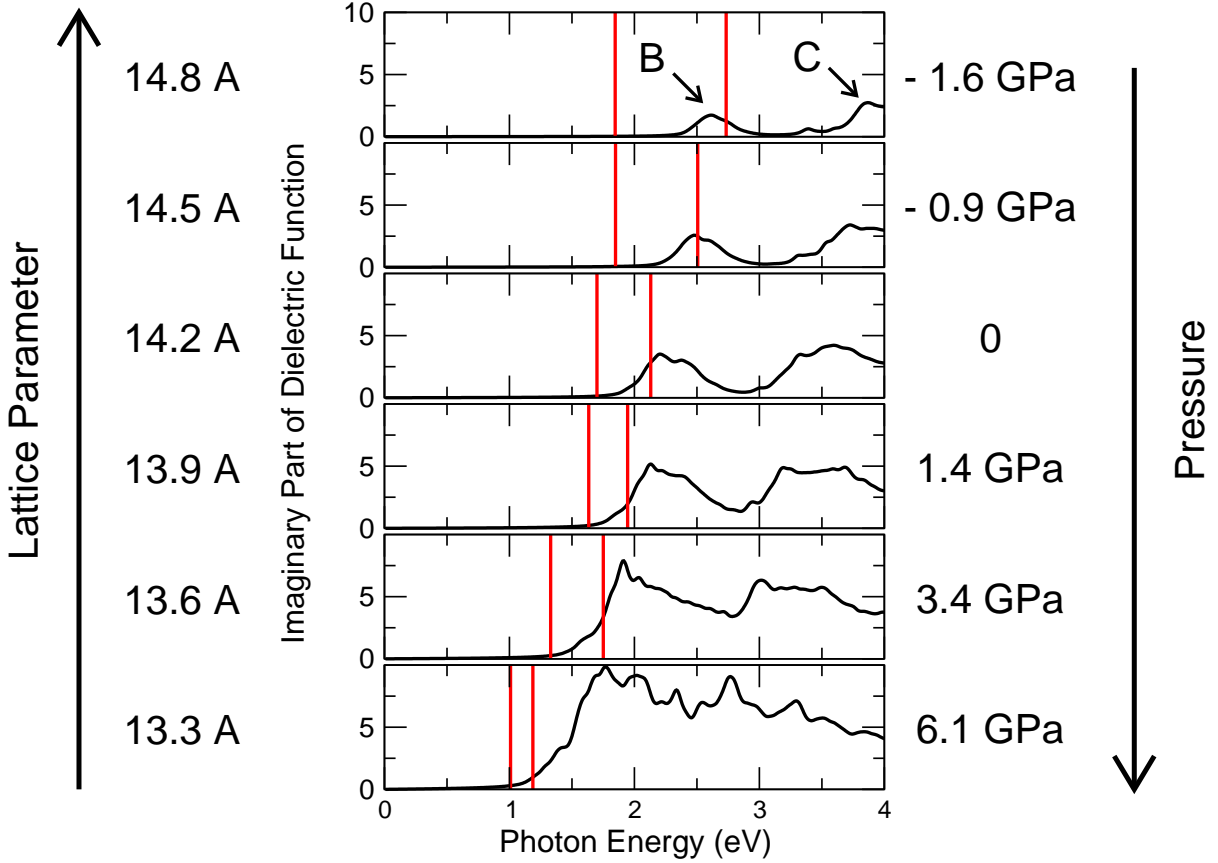}
\caption{(Color online) Imaginary part of the dielectric function for several choices
of lattice parameter. Vertical bars on each panel indicate the
calculated optical and electronic gaps. An artificial Gaussian
broadening of 0.02~eV was added to all absorption spectra. Sharp
features in the spectrum are expected to fade away with inclusion of
rotational disorder. Line A in the measured spectrum \cite{MeletovD98}
(see text) is very weak to be visible. Line B is the first absorption
line, at around 2.2~eV at zero pressure. Line C is the second
absorption line, at around 3.6~eV at zero pressure.}
\label{fig.spectrum}
\end{figure}

Figure \ref{fig.gap} shows that the electronic and optical gaps
decrease continuously as an hydrostatic pressure of up to 6~GPa is
applied on crystalline C$_{60}$. The overall decrease in electronic
gap is 0.9~eV with pressure ranging from zero to 6~GPa. To our
knowledge, the electronic gap at high hydrostatic pressure has not
been measured yet. The optical gap ({\em i.e.}, the excitation energy of
the first bound exciton) decreases by 0.7~eV in the same pressure
range. Since that exciton is optically inactive, the best comparison of
optical activity as a function of pressure should be done 
following the position of the first peak in the absorption spectrum,
on Figure \ref{fig.spectrum}. The peak moves from 2.2~eV to 1.75~eV in
the pressure range from zero to 6~GPa. This is compatible with the
first determinations of transmission edge as a function of pressure
\cite{SnokeSM93}: the transmission edge decreases from 1.9~eV (zero
pressure) to 1.5~eV (5~GPa).

Figure \ref{fig.gap} also shows that the
profile of energy gap versus lattice parameter is not linear. A
suitable model for the dependence of the gap with respect to pressure
should take into account the behavior
of dielectric screening for different amounts of intermolecular
spacing and hence different amounts of overlap between molecular
orbitals at different molecules. Snoke and collaborators
\cite{SnokeSM93} have proposed a phenomenological model for the gap.

\begin{figure}
\includegraphics[width=16cm]{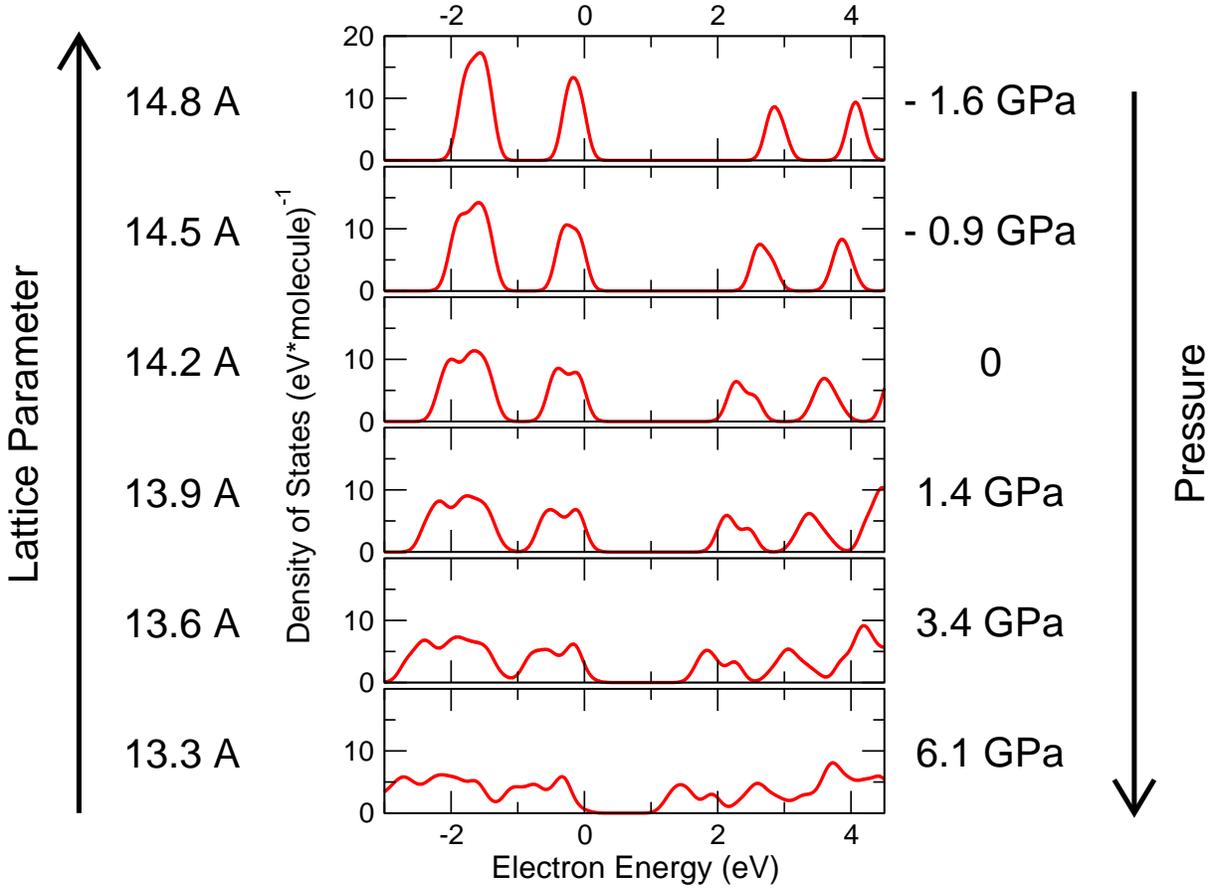}
\caption{(Color online) Density of states in fullerite for several choices of lattice
parameter, obtained within the GW theory. Energies are defined with
respect to the valence band maximum. A Gaussian broadening of 0.1~eV
was added to all density distributions. The five major features in the
density of states (well separated at zero and ``negative'' pressure)
correspond to different molecular multiplets, from lower to higher
energy: $H_g$+$G_g$ (superimposed) , $H_u$, $T_{1u}$, $T_{1g}$.}
\label{fig.dos}
\end{figure}

Meletov and Dolganov \cite{MeletovD98} have also found a decrease in
the optical gap as a function of pressure. In their experiment,
microcrystals of fullerite were placed inside a diamond anvil cell,
with pressure of up to 2.5~GPa. Several phenomena were observed in
that experiment:

\begin{enumerate}
\item At zero pressure, a low-energy line and two well pronounced
  lines in the absorption spectrum were found, labeled A (at 2.0~eV),
  B (2.7~eV) and C (3.5~eV) respectively. Line A is very weak and it
  could originate from transitions $H_u \to T_{1u}$, which gain finite
  oscillator strength from mixing with higher transitions. That
  interpretation is supported by our calculations, which indicate an
  onset of the line at 1.7~eV and very small but non-vanishing
  oscillator strength.

\item Lines B and C have similar strength. It was found experimentally 
  that optical activity  migrates from C to B as the microcrystals are
  compressed. That effect is found in Figure \ref{fig.spectrum}, where
  we see enhancement of the peak at 2.2~eV and reduction of the peak
  at 3.5~eV, while both peaks redshift from zero to 3.4~GPa. Since we
  also see mixing between transitions $H_u \to T_{1g}$ and $H_g \to
  T_{1u}$, the major components of peaks B and C respectively, our
  calculations confirm the assumption that migration of optical
  activity is caused by mixing between different optical transitions \cite{MeletovD98}.
  
\item The energy dependence of the measured absorption spectrum was
  reported to be weakly dependent on pressure in the pressure range
  from zero to 2.5~GPa \cite{MeletovD98}. Figure \ref{fig.spectrum}
  confirms that observation. At the next pressure value (6.1~GPa), the
  two peaks merge into an asymmetric wide peak. That indicates that
  bands derived from different molecular multiplets start to overlap,
  as shown in Figure \ref{fig.dos}.
\end{enumerate}

Hydrostatic pressure also modifies the character of bound excitons,
making them more delocalized. Comparing the distribution of
probabilities at zero pressure (lattice parameter $14.2~\AA$) and
3.4~GPa (lattice parameter $13.6~\AA$), Table \ref{table.probability}
shows that the first bound exciton becomes substantially more
delocalized, with a correlation radius between the first and third
nearest-neighbor distances. We believe that two mechanisms contribute
to delocalization: applied pressure increases the overlap of molecular
orbitals on different molecules, thus increasing the probability of
one electron moving from one molecule to its neighbor; and pressure
also increases the mixing between bands, particularly between
transition $H_u \to T_{1u}$ and higher transitions.

As in the zero-pressure regime, we use the $Pa{\overline{3}}$ lattice
to perform calculations at hydrostatic pressure, with no orientational
disorder. Our estimates of the impact of disorder on the energy gap,
mentioned at the end of the previous section, also apply to the regime
of finite pressure. Since our calculations do not contain accurate van
der Waals forces, we have not attempted to investigate the emergence
of different glassy phases as a function of pressure. Including
accurate van der Waals forces would remove the inaccuracy of the
calculated gaps with respect to orientational disorder in fullerite.

\section{Fullerite with intercalated molecules} \label{results_cubane}

Fullerene C$_{60}$ is very stable, which favors the engineering of
microcrystals with intercalated molecules. At equilibrium, the FCC
crystal has two large types of voids: an octahedral site with radius
3.5~\AA \  and a tetrahedral site with radius 1.15~\AA. Isolated atoms
and small molecules can be easily placed in one of those voids. Doped
C$_{60}$ has very interesting properties, for instance K$_3$C$_{60}$
is superconductor at 18~K \cite{HebardRHMGPRK91}. Ca$_3$C$_{60}$ is
superconductor at 8.4~K \cite{KortanKGGRFTH92}. Those compounds also
show significant electron transfer from dopant atom to cage. Doping
fullerite with wide-gap molecules produce different
phenomena. Depending on the concentration and symmetry of the dopant, it
can lower the symmetry of the host crystal and enhance the oscillator
strength of otherwise dark optical transitions of
fullerite. Highly-symmetric dopants are expected to produce less
distortions in the host. In particular, cubane (C$_8$H$_8$) has been
proposed as an ideal intercalator \cite{PekkerKOBKBJJBKBKTF05}. It has perfect cubic symmetry. If
placed in an octahedral void, it will preserve the cubic symmetry of
the lattice. With doping, the crystal is forced to expand
isotropically in order to accommodate the extra molecules but no
additional structural distortion is necessary. In addition, solid
cubane is bound by weak van der Waals forces \cite{YildirimGNEE92}, which
means that cubane is not likely to segregate into clusters. The
ionization potential of molecular cubane is 8.6~eV
\cite{LifshitzE83}. Its electron affinity is negative, indicating that
it has an energy gap in the ultraviolet range. Since the band edges of
C$_{60}$ are inside the ones of cubane, cubane can be used to mimic negative hydrostatic pressure in fullerite
without altering the optical properties of the host \cite{PekkerKOBKBJJBKBKTF05}.

\begin{figure}
\includegraphics[width=16cm]{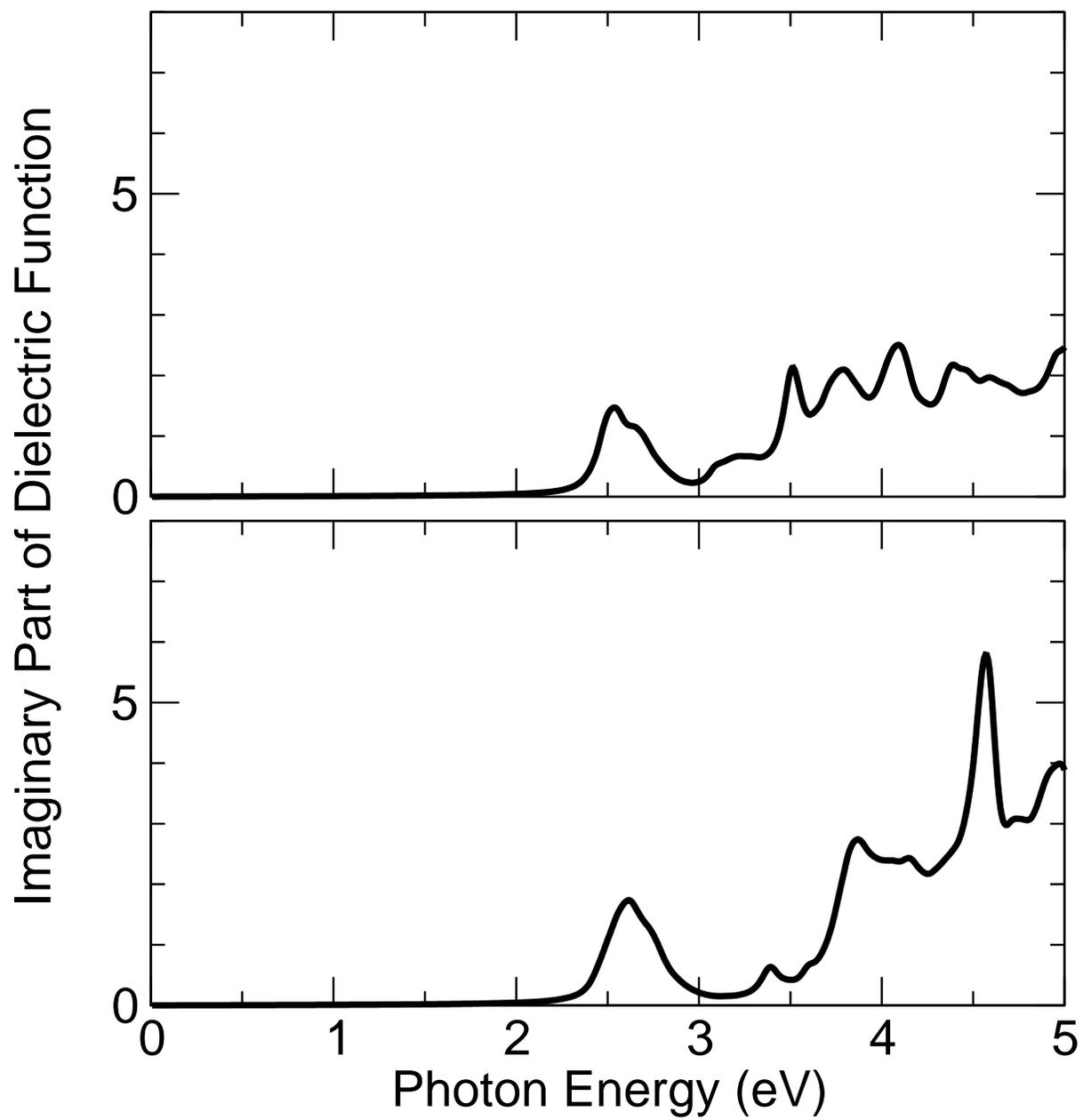}
\caption{Imaginary part of the dielectric function of
  C$_8$H$_8$-C$_{60}$ (a) and pure C$_{60}$ (b). An
  artificial Gaussian broadening of 0.02~eV was added to all absorption
  spectra.}
\label{fig.cubane_spectrum}
\end{figure}

We built a lattice of cubane-fullerene with maximum doping by filling
all octahedral voids in the FCC lattice with cubane. The lattice
parameter is taken as 14.8~\AA, following experimental determination
of the rotor-stator phase \cite{PekkerKOBKBJJBKBKTF05}. The band
structure of this compound around its energy gap is very similar to
undoped fullerite with the same lattice constant. Cubane derived bands
are found no less than 4~eV away from the gap. The electronic gap of
fullerite with intercalated cubane (C$_8$H$_8$-C$_{60}$), obtained
from our GW calculations is 2.6~eV, similar to the electronic gap
obtained for fullerite with the same lattice parameter (2.7~eV). The
difference of 0.1~eV is close to the numerical precision of our
calculations. C$_8$H$_8$-C$_{60}$ and pristine fullerite also have
similar optical gaps: 2.0~eV and 1.85~eV. Figure
\ref{fig.cubane_spectrum} shows that the absorption spectra of
C$_8$H$_8$-C$_{60}$ and pristine fullerite differ from each other only
above 3.5~eV. These findings confirm that the effect of adding cubane
is, by and large expand the lattice of fullerene molecules. All other
phenomena in its electronic structure are direct consequences of
lattice expansion.

Other intercalants can also produce lattice expansion. Atoms of noble
gases are good candidates, owing to their low reactivity and high
energy gap. One shortcoming is that they are smaller than
cubane. While these molecules will not expand the lattice
significantly, a small change
in gap can be measured. Therefore, fullerite could be used as a
sensor of inert molecules. Significant
expansion could be obtained by overdoping fullerite with several atoms
per interstitial site. Other candidates are small molecules such as
methane (CH$_4$), hydrogen (H$_2$) or  nitrogen (N$_2$).

\section{Summary and perspectives} \label{conclusion}

The results presented above show that several properties of fullerite,
particularly the threshold of its optical absorption, can be tuned by
applying pressure. By applying hydrostatic pressure of up to 6~GPa,
easily obtained in diamond anvil cell devices, the first peak of
optical absorption redshifts from 2.2~eV (yellow-green) to 1.8~eV
(red), thus making microcrystals less transparent. Similar reduction
in the gap as a function of pressure has been reported in alkali-doped
fullerite \cite{PoloniFFPTMCPS08}. Applications of
this phenomenon are plenty. One of them is in piezo-optical sensors:
one can put clean microcrystals of fullerite in an environment under
unknown pressure and infer the pressure by measuring their
transmittance or absorbance. This is particularly useful if the
microcrystals are part of a microdevice, subject to pressure gradients
and where usual pressure gauges cannot be used.

One can expand the range of colors where fullerite gauges operate by
doping microcrystals with weakly interacting molecules. It has been
shown experimentally that saturating microcrystals with cubane
increases its lattice parameter \cite{KortanKGGRFTH92}. Our results
show that the lattice expansion produces a blueshift of the first
absorption peak from 2.2~eV to around 2.6~eV. Other dopants can
produce larger lattice expansion and hence larger blueshifts,
depending on their size and concentration.

In summary, we have done first-principles calculations of electronic
and optical properties of fullerite in order to characterize their
pressure dependence. Comparison between available experimental data
and our calculations at equilibrium lattice parameter show that our
methodology predicts gaps with an accuracy of 0.1 to 0.2~eV. The
absorption edge shifts toward the red end of the spectrum as we apply
hydrostatic pressure of up to 6~GPa. There is little distortion in the
electronic structure of the material in the pressure range
investigated. We have also confirmed earlier hypotheses that
cubane-intercalated fullerite has optical properties very similar to
fullerite with an artificial lattice expansion. These findings show
that pure fullerite or fullerite with inert dopants can be used as
active element in piezo-optical sensors.

We would like to thank discussions with E. Schwegler, T. Oguitsu and 
H. Whitley for discussions. 
Research sponsored by the Division of Materials Sciences and
Engineering BES, U.S. DOE under contract with UT-Battelle,
LLC. Computational support was provided by the National Energy
Research Scientific Computing Center.


\begin{thebibliography}{30}
\bibitem{KrotoHOCS85} H. Kroto, J. Heath, S.~O'Brien, R. Curl, and R. Smalley, Nature {\bf 318}, 162 (1985).

\bibitem{DresselhausDE96} M.~S. Dresselhaus, G. Dresselhaus, and P.~C. Eklund, \emph{Science of fullerenes and carbon nanotubes: their properties and applications} (Academic Press, San Diego, 1996).

\bibitem{KratschmerLFH90} W. Kratschmer, L. Lamb, K. Fostiropoulos, and D. Huffman, Nature {\bf 347}, 354 (1990).

\bibitem{BlaseGSM04} X. Blase, P. Gillet, A. San Miguel, and P. M\'elinon, Phys. Rev. Lett. {\bf 92}, 215505 (2004).

\bibitem{San-Miguel06} A. San Miguel, Chem. Soc. Rev. {\bf 35}, 876 (2006).

\bibitem{KortanKGGRFTH92} A. Kortan, N. Kopylov, S. Glarum, E. Gyorgy, A. Ramirez, R. Fleming, F. Thiel, and R. Haddon, Nature {\bf 355}, 529 (1992).

\bibitem{HebardRHMGPRK91} A. Hebard, M. Rosseinsky, R. Haddon, D. Murphy, S. Glarum, T. Palstra, A. Ramirez, and A. Kortan, Nature {\bf 350}, 600 (1991).

\bibitem{MeletovK05} K. Meletov and G. Kourouklis, J. Exp. Theor. Phys. {\bf 100}, 760 (2005).

\bibitem{Sundqvist99} B. Sundqvist, Adv. Phys. {\bf 48}, 1 (1999).

\bibitem{MeletovD98} K. Meletov and V. Dolganov, J. Exp. Theor. Phys. {\bf 86}, 177 (1998).

\bibitem{PekkerKOBKBJJBKBKTF05} S. Pekker, E. Kovats, G. Oszlanyi, G. Benyei, G. Klupp, G. Bortel, I. Jalsovszky, E. Jakab, F. Borondics, K. Kamaras, et al., Nature Mat. {\bf 4}, 764 (2005).

\bibitem{PoloniFFPTMCPS08} R. Poloni, M. V. Fernandez-Serra, S. Le Floch, S. De Panfilis, P. Toulemonde, D. Machon, W. Crichton, S. Pascarelli, and A. San-Miguel, Phys. Rev. B {\bf 77}, 035429 (2008).

\bibitem{Duclos91} S. Duclos, K. Brister, R. Haddon, A. Kortan, and F. Thiel, Nature {\bf 351}, 380 (1991).

\bibitem{NunezMHBP95} M. N\'u\~nez Regueiro, L. Marques, J.~L. Hodeau, O. B\'ethoux, and M. Perroux, Phys. Rev. Lett. {\bf 74}, 278 (1995).

\bibitem{SnokeSM93} D.~W. Snoke, K. Syassen, and A. Mittelbach, Phys. Rev. B {\bf 47}, 4146 (1993).

\bibitem{MosharyCSBDVB92} F. Moshary, N.~H. Chen, I.~F. Silvera, C.~A. Brown, H.~C. Dorn, M.~S. de Vries, and D.~S. Bethune, Phys. Rev. Lett. {\bf 69}, 466 (1992).

\bibitem{KozlovY95} M. Kozlov and K. Yakushi, J. Physics - Condens. Matter {\bf 7}, L209 (1995).

\bibitem{ShirleyL93} E.~L. Shirley and S.~G. Louie, Phys. Rev. Lett. {\bf 71}, 133 (1993).

\bibitem{ShirleyBL96} E.~L. Shirley, L.~X. Benedict, and S.~G. Louie, Phys. Rev. B {\bf 54}, 10970 (1996).

\bibitem{HartmannZMSBLFZ95} C. Hartmann, M. Zigone, G. Martinez, E.~L. Shirley, L.~X. Benedict, S.~G. Louie, M.~S. Fuhrer, and A. Zettl, Phys. Rev. B {\bf 52}, R5550 (1995).

\bibitem{Martin} R.~W. Martin, \emph{Electronic structure: basic theory and practical methods} (Cambridge University Press, Cambridge, UK, 2004).

\bibitem{PerdewBE96} J.~P. Perdew, K. Burke, and M. Ernzerhof, Phys. Rev. Lett {\bf 77}, 3865 (1996).

\bibitem{HybertsenL86} M.~S. Hybertsen and S.~G. Louie, Phys. Rev. B {\bf 34}, 5390 (1986).

\bibitem{AulburJW00} W. Aulbur, L. J\"onsson, and J. Wilkins, \emph{Solid State Physics} (Academic Press, New York, 2000), vol. 54, p. 1.

\bibitem{OnidaRR02} G. Onida, L. Reining, and A. Rubio, Rev. Mod. Phys. {\bf 74}, 601 (2002).

\bibitem{RohlfingL00} M. Rohlfing and S.~G. Louie, Phys. Rev. B {\bf 62}, 4927 (2000).

\bibitem{FetterW} A.~L. Fetter and J.~D. Walecka, \emph{Quantum Theory of Many-Particle Systems} (Mc-Graw Hill, New York, 1971).

\bibitem{TiagoC06} M.~L. Tiago and J.~R. Chelikowsky, Phys. Rev. B {\bf 73}, 205334 (2006).

\bibitem{SpataruICL05} C.~D. Spataru, S. Ismail-Beigi, R.~B. Capaz, and S.~G. Louie, Phys. Rev. Lett. {\bf 95}, 247402 (2005).

\bibitem{TiagoKHR08} M.~L. Tiago, P.~R.~C. Kent, R.~Q. Hood, and F.~A. Reboredo, J. Chem. Phys. {\bf 129}, 084311 (2008).

\bibitem{Cotton} F.~A. Cotton, \emph{Chemical Applications of Group Theory} (J. Wiley and Sons, New York, 1990).

\bibitem{YildirimGNEE92} T. Yildirim, P.~M. Gehring, D.~A. Neumann, P.~E. Eaton, and T. Emrick, Phys. Rev. Lett. {\bf 78}, 4938 (1997).

\bibitem{LifshitzE83} C. Lifshitz and P. Eaton, Int. J. Mass Spectrom. Ion Phys. {\bf 49}, 337 (1983).

\end{thebibliography}

\newpage

\newpage

\newpage

\newpage

\newpage

\begin{table}
\begin{ruledtabular}
\begin{tabular}{cccc}
Location of Electron &
\multicolumn{3}{c}{Probability} \\
& $a = 13.6 \AA$ & $a = 14.2 \AA$ & $a = 14.8 \AA$ \\ \hline
Hole's molecule & 15 \% & 62 \% & 84 \% \\
1$^{st}$ Nearest Neighbor & 25 \% & 30 \% & 13 \% \\
2$^{nd}$ Nearest Neighbor & 10 \% & 2 \% & $<$ 1\% \\
3$^{rd}$ Nearest Neighbor & 25 \% & 3 \% & $<$ 1 \% \\
\end{tabular}
\end{ruledtabular}
\caption{ Electron probability distribution of  the first bound 
exciton calculated as a function of the distance to the hole. 
The probability was  integrated over each shell of molecules 
around the molecule that contains the hole. The integration was
performed over a Wigner-Seitz cell centered on each molecule. Three
different lattice parameters are shown: 13.6~$\AA$ (3.4~GPa),
14.2~$\AA$ (zero pressure) and 14.8~$\AA$ (negative pressure
 -1.6~GPa).}
\label{table.probability}
\end{table}

\end{document}